# Correlations of neutron multiplicity and gamma multiplicity with fragment mass and total kinetic energy in spontaneous fission of $^{252}$Cf

Taofeng Wang[1,2], Guangwu Li[3], Wenhui Zhang[3], Liping Zhu[3], Qinghua Meng[3], Liming Wang[3], Hongyin Han[3], Haihong Xia[3], Long Hou[3], Ramona Vogt[4,5], and Jørgen Randrup[6]

[1]*School of Physics and Nuclear Energy Engineering, Beihang University, Beijing 100191, China*
[2]*International Research Center for Nuclei and Particles in the Cosmos, Beijing 100191, China*
[3]*China Institute of Atomic Energy, P.O. Box 275 - 46, Beijing 102413, China*
[4]*Nuclear and Chemical Sciences Division, Lawrence Livermore National Laboratory, Livermore, CA 94551, USA*
[5]*Physics Department, University of California at Davis, Davis, CA 95616, USA*
[6]*Nuclear Science Division, Lawrence Berkeley National Laboratory, Berkeley, CA 94720, USA*

**Abstract**

The dependence of correlations of neutron multiplicity $v$ and gamma-ray multiplicity $M_\gamma$ in spontaneous fission of $^{252}$Cf on fragment mass $A^*$ and total kinetic energy TKE has been investigated employing the ratio of $M_\gamma/v$ and the form of $M_\gamma(v)$. We show for the first time that $M_\gamma$ and $v$ have a complex correlation for heavy fragment masses, while there is a positive dependence of $M_\gamma(v)$ for light fragment masses and for near-symmetric mass splits. The ratio $M_\gamma/v$ exhibits strong shell effects for the neutron magic number $N = 50$ and near the doubly magic number shell closure at $Z = 50$ and $N = 82$. The gamma-ray multiplicity $M_\gamma$ has a maximum for TKE = 165-170 MeV. Above 170 MeV $M_\gamma$(TKE) is approximately linear, while it deviates significantly from a linear dependence at lower TKE. The correlation between the average neutron and gamma-ray multiplicities can be partly reproduced by model calculations.

PACS numbers: 24.75.+i; 25.85.Ca; 25.85.Ec; 25.85.Ge

## 1. Introduction

Nuclear fission is a complicated dynamical process in which a heavy nucleus develops into two excited and distorted pre-fragments at scission. Because of the excitation of the dinuclear bending and wriggling modes, as well as the Coulomb torque between two fragments after scission, the fragments emerge with significant angular momenta. As the emerging fragments relax to their equilibrium shapes, the potential energy associated with their initial shape distortion is converted into additional statistical excitation energy. Each fragment subsequently disposes of its excitation energy by neutron evaporation and, later on, by gamma radiation. Consequently, the multiplicities of the promptly emitted neutrons and gamma-rays are intimately related to the initial fragment excitation energy and the initial fragment angular momenta.

In spontaneous fission of $^{252}$Cf, the neutron multiplicity exhibits a familiar sawtooth structure as a function of the fragment mass [1-5]. The total kinetic energy (TKE) is high at $A^* = 132$ [6, 7] and the neutron emission is reduced, because the heavy fragment in this region is close to the doubly magic $^{132}$Sn nucleus. Because the heavy pre-fragment is then closer to sphericity, the Coulomb repulsion at scission is larger, resulting in a larger relative kinetic energy. Consequently, the fragment excitation energy is lower, causing the neutron and/or gamma-ray emission from the fragments to be reduced. Conversely, the low-TKE fission mode arises from very elongated scission shapes [8] consisting of highly deformed pre-fragments.

The prompt gamma-rays are primarily emitted by both statistical and collective de-excitation of the fission fragments [9], after they have cooled down through neutron evaporation. The gamma-ray yield as a function of the mass split is very sensitive to the initial sharing of excitation energy among the two fragments and to their level-density parameters [10]. One can investigate simultaneously gamma-ray emission from nuclei with masses near shell closures as well as in well-deformed and soft deformable regions, where gamma-ray emission is governed by distinctly different mechanisms. The dependence of the gamma-ray multiplicity $M_\gamma$ on fragment mass is not fully understood. Previous measurements have reported a sawtooth-like behavior of $M_\gamma$ for $^{252}$Cf(sf) [11] and $^{235}$U($n_{th}$,f) [12], similar to that of the neutron multiplicity, whereas Glässel et al. [13] found $M_\gamma$ to be rather independent of fragment mass.

It is well known that both the average neutron multiplicity $v$ and the average radiated gamma-ray energy $E_\gamma$ increase with excitation energy [14], suggesting a positive correlation of $v$ and $E_\gamma$. Nifenecker et al. [15, 16] suggested the relation $E_\gamma = 0.75v + 2.0$ by correlating $v$ and $E_\gamma$ of individual fragments for $^{252}$Cf(sf). Since determining the dependence of the gamma-ray yield on the fragment mass is difficult, the above empirical formula is typically employed in model calculations and nuclear data evaluations. However, several explorations [13, 17] in recent years indicate that this correlation is likely not accurate. To clarify this and provide new information on the fission mechanism, we investigated the correlation between gamma-ray multiplicity $M_\gamma$ and neutron multiplicity $v$ by measuring the ratio $M_\gamma/v$ as a function of fragment mass $A^*$ as well as the dependence of $M_\gamma$ on $v$ in several fragment mass regions for $^{252}$Cf(sf). Moreover, the dependence of the average gamma-ray multiplicity on TKE was also determined.

There are a number of complete event Monte Carlo models of fission, including CGMF [21], FIFRELIN [20], FREYA [18, 19], and GEF [22]. All these models emit neutrons from fragments down to the neutron separation energy, followed by gamma emission. Since all emitted neutrons and gammas can be tracked and associated with a specific fragment, neutron and gamma emission can be correlated. CGMF and FIFRELIN employ a Hauser-Feshbach framework for fragment de-excitation. FREYA and GEF model neutron emission with a Weisskopf-Ewing spectral shape. CGMF, FIFRELIN, and FREYA employ data-driven fragment yields and total kinetic energy distributions as inputs, while GEF employs yields determined by the potential energy landscape between the fission barrier and scission as a function of the mass asymmetry. Instead of using TKE as an input, the excitation energy in GEF is calculated and partitioned between the light and heavy fragments by a probability distribution based on the product of the fragment level densities.

We compare our measurements to FREYA and GEF calculations. In both codes, neither the neutron evaporation nor the subsequent statistical dipole gamma emission changes the fragment angular momentum substantially, so the fragment rotational energy is primarily disposed of by E2 transitions along the yrast line. FREYA explicitly conserves angular momentum. It assumes that, at scission, the angular momenta of the fragments are perpendicular to the line joining the dinuclear axis. Thus bending and wriggling modes are included, while tilting and twisting modes are ignored. Because fragment deformation is not explicitly included, as in GEF, a parameter is employed to redistribute the excitation energy between the light and heavy fragments.

## 2. Experimental procedure

For experimental fission physics studies, it is generally necessary to manufacture a thin and symmetrical source of fissional material on an easily penetrable foil. Because the atoms of $^{252}$Cf exhibit the self-transfer characteristics due to the kinetic energy transfer from fission fragments to the atoms, the atom agglomerates consisting of thousands of $^{252}$Cf atoms would emit from the surface of the mother source under vacuum conditions. The self-transfer rate of the $^{252}$Cf atoms will mainly depend on the purity and thickness of the mother $^{252}$Cf source as well as the vacuum level of the chamber. The vacuum level of the chamber was kept as ~ $10^{-2}$ mm Hg during manufacture of the spontaneous fission foil $^{252}$Cf source. A metal collimator with a 5 mm diameter circular hole was placed on the mother source, 5 mm away from the thin carbon foil with a thickness of 40 μg/cm$^2$. This thin foil backing, mounted on a copper ring of 0.5 mm thickness, 28 mm outer diameter, and 16 mm inner diameter, was fixed on a stand. During the experiment, the fission source was mounted between two silicon surface barrier detectors (F1 and F2) face-to-face, which were employed to measure the fission fragment kinetic energies. The schematic of the experimental setup is shown in Fig. 1. The distances from the fission source to F1 and F2 were 6 cm and 4 cm, respectively. Both surface barrier detectors (diameter Φ = 20 mm) were collimated down to 16 mm in diameter to avoid edge effects. The fission source and the fission detectors were placed in a cylindrical copper chamber (Φ = 30 cm × 25 cm) with a wall thickness of 2 mm and at a vacuum of about 0.2 Pa. A cylindrical liquid scintillator (Φ = 10 cm × 5 cm) held with wire 46 cm behind F1 served as neutron detector, while a HPGe detector behind the F2, with efficiency 60% relative to the NaI(Tl) gamma-ray detector, was placed 36 cm from F2. The HPGe detector was shielded by lead bricks. To reduce the effect of neutron scattering from the wall and other materials, the experimental hall was large and as empty possible. Separate data are taken with the HPGe detector in different locations, namely, at 0 and 90 degrees. In the case of 0 degrees, all detectors (F1, F2, liquid scintillator, HPGe) and the fission source are placed coaxially. In the 90 degree configuration, the HPGe detector is perpendicular to the axis of the other detectors. To determine the absolute values of $v(A^*)$ and $M_\gamma(A^*)$ for $85 < A^* < 167$, the kinematical focus effect of neutron emission from moving fragments and the Doppler shift effect of gamma-rays from the moving emission source were employed.

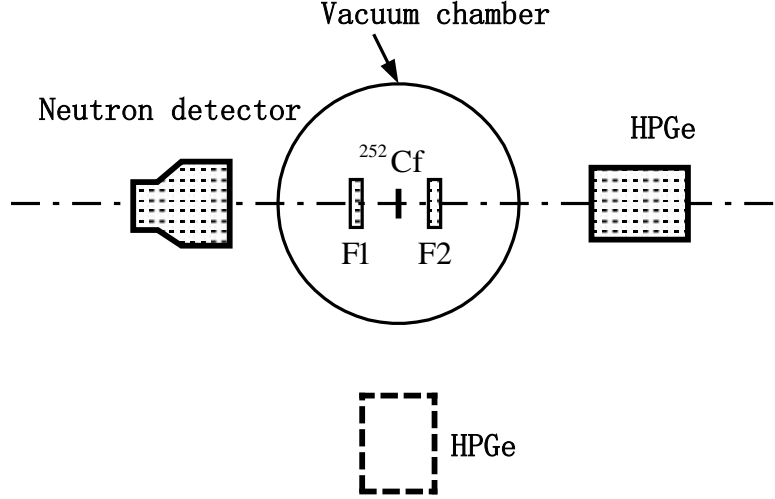

Fig.1 Schematic diagram of the experimental setup.

The energy calibration of the silicon surface barrier detectors is not as straightforward as it is for light ions such as alpha particles. The main reasons are the existence of pulse height defects due to nuclear collisions, incomplete collection of charge, and energy loss in the dead layer (window). The fragment kinetic energies were obtained from the F1 and F2 pulse heights of by using the Schmitt formula [23].

$$E_i = (a + a'A)X + b + b'A, \qquad (1)$$

where $E_i$ is the kinetic energy of the fragment $F_i$ ($i=L, H$), $X$ is the pulse height, and $A$ is the mass. The coefficients $a$, $a'$, $b$, and $b'$ are determined from the locations $P_L$ and $P_H$ of the light and heavy fragment pulse height peaks as [24].

$$a = a_0/(P_L - P_H); a' = a_0'/(P_L - P_H);$$

$$b = b_0 - aP_L; b' = b_0' - a'P_L, \qquad (2)$$

where $a_0$, $a_0'$, $b_0$, and $b_0'$ are universal constants given in Ref. [24]. The pre-neutron-emission kinetic energies $E_i^*$ are expressed as the following equations:

$$E_L^* = E_L + E_{nL}, \qquad (3)$$

$$E_H^* = E_H + E_{loss} + E_{nH}, \qquad (4)$$

where $E_{ni}$ refers to the kinetic energy carried away by neutrons emitted from the fragment $i=L, H$ and $E_{loss}$ is the energy loss of the fragment in the $^{252}$Cf source backing. On the basis of momentum and mass conservation, an iterative method was employed to solve Eqs. (3) and (4) for each measured fission event. At the beginning of the iteration, it is assumed that the quantities $E_{ni}$ and $E_{loss}$ in Eqs. (3) and (4) together with $A$ in Eq. (1) are equal to be zero. Therefore the provisional fragment mass $A_i^*$ and provisional fragment kinetic energy $E_i^*$, before the neutron emission, are derived. Thus, $A_L^* = A_0 E_H^* / TKE$, $A_H^* = A_0 - A_L^*$ and the total kinetic energy of fragments $TKE = E_L^* + E_H^*$, where $A_0$ is the mass of the compound nucleus $^{252}$Cf. Using the $\bar{\nu}$ values given by Ref. [5], the post-neutron-emission fragment mass $M_i$, and the quantities $E_{ni}$ are given by the following relations,

$$A_i = A_i^* - \bar{\nu}(A_i^*, TKE), \tag{5}$$

$$E_{ni} = \frac{\bar{\nu}(A_i^*, TKE) E_i^*}{A_i^*}, \tag{6}$$

The values of $A_i$ and $E_{ni}$ obtained from Eqs. (5) and (6) are adopted for the calculation of the value of $E_{loss}$ [25]. The extracted values of $E_{ni}$, $E_{loss}$ and $A$, are inserted into Eqs. (1)-(4) and new values of $A_i^*$ and $E_i^*$ are found. With these new values, updated values of $A_i$, $E_{ni}$, and $E_{loss}$ are obtained from Eqs. (5) and (6). When the differences between energies calculated in two consecutive iterations are less than 100 keV, the pre-neutron emission fragment mass $A_i^*$ and kinetic energy $E_i^*$ are accepted. A TKE interval of 3 MeV, which corresponds to the fragment energy resolution for silicon barrier detectors, is adopted. The mass resolution of the detection system is estimated as 4.1 u. Experimental systematic uncertainties are hard to assess and the disentanglement of uncertainty correlations can be a complex procedure.

The geometrical detection efficiency of the LS301 neutron detector was calculated with the Monte Carlo code NEFF50 [26], using the measured light output function. The calculated efficiencies were corrected by comparing the measured pulse height spectra with the calculated spectra. In this way, the efficiency was obtained to an accuracy of about 3% for energies from 2 to 15 MeV. The calculated efficiencies below 6 MeV for different thresholds were checked using a mini fast ionization chamber combined with a $^{252}$Cf source [27] which has a standard fission neutron spectrum. The comparison between the measurements and calculations are shown in Fig. 2. The mini ionization chamber provided the neutron time of flight (TOF) start signal for fission fragment detection. The stop signal was given by the anode of neutron detector. The measured TOF spectra were converted to the neutron energy spectra and compared to the standard. Background subtraction was carried out. The $^{252}$Cf neutron spectrum was modified relative to a Maxwellian with temperature of 1.42 MeV. Neutron scattering corrections were made to the fission neutron spectrum. The fraction of scattered neutrons contaminating the source neutron spectrum in the 0.9 g ionization chamber depends on the neutron energy [28]: ~1.5% for neutron energies below 0.5 MeV; ~1.2% for neutron energies from 0.5 to 1.0 MeV; ~0.7% for neutron energies from 1.0 to 2.0 MeV; and less than 0.5% for the neutron energies above 2 MeV. The resulting neutron detection efficiency is in good agreement with the calculated efficiencies for energies below 6 MeV [27]. The efficiency uncertainty for energies above 6 MeV is larger than 10% due to low statistics of high energy fission neutrons.

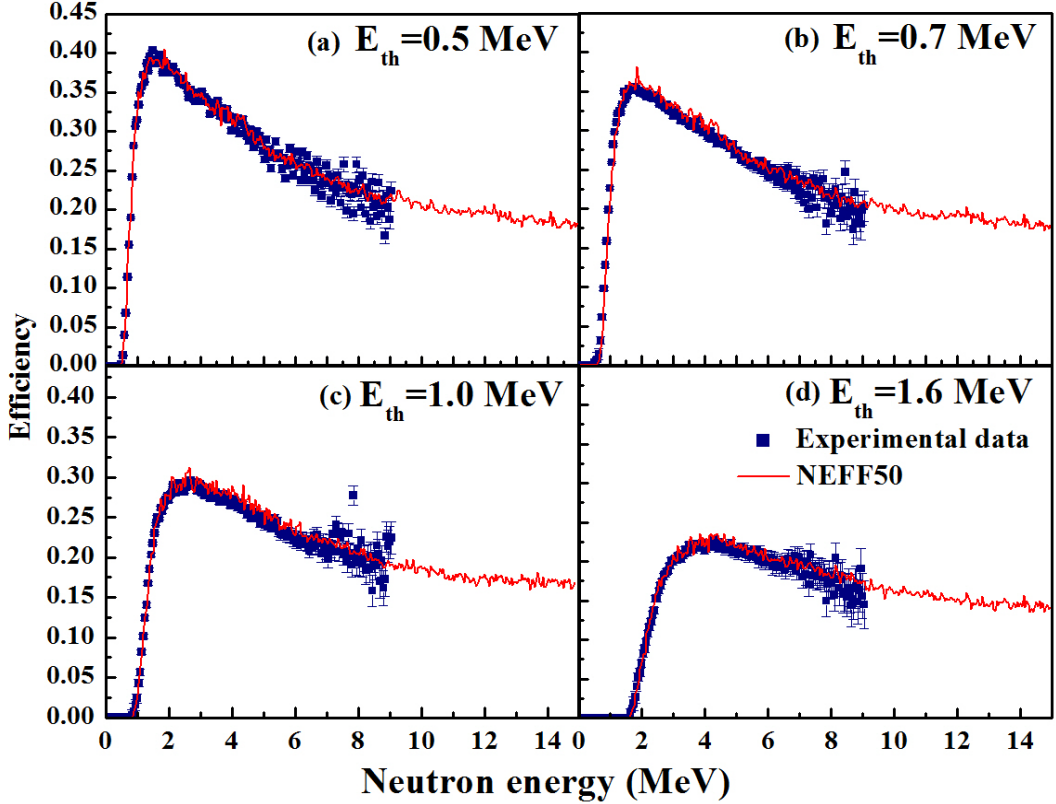

Fig. 2 The efficiency of the LS301 neutron detector for different energy thresholds.

Misidentification of neutrons from the complementary fragment was estimated using a Monte Carlo simulation that included the fragment velocity, the neutron spectrum in the center of mass system, and the scintillator threshold energy. The contamination from misidentified neutrons was found to be a few percent, even for the strictest condition. Therefore this effect was neglected. The neutron multiplicity was determined by comparing the number of coincident fission neutrons to the number of fissions, obtained from the detection of F1 or F2. Corrections due to neutron and gamma-ray scattering and absorption were carried out during the data analysis.

The assignment of individual fragment gamma-ray multiplicities $M_\gamma$ were made by employing the Doppler shift method [29] which exploits the shift of the gamma-ray energies and/or the change of the angular distribution due to the fragment motion. The experimental data of the 0° HPGe detection involving the gamma-rays with the Doppler effect were used to determine the absolute gamma-ray multiplicity value for fragment mass $A^*$. The data measured in the 90° HPGe detection without the Doppler effect were used to determine the correlation of the total gamma-ray multiplicity value with fragment mass. The average total number of gamma-rays emitted per fission is given by [12]

$$\omega = [\frac{n_{\gamma A}}{p_\gamma} + \frac{n_{\gamma(A_0-A)}}{p_\gamma}]/2n_A, \qquad (7)$$

where $n_{\gamma A}$ and $n_{\gamma(A_0-A)}$ are the number of counts observed in the gamma-ray detector when a fragment of mass $A$ is detected by F1 and a fragment of mass $A_0$ - $A$ is detected by F2. $n_A$ is the total number of fissions for which fragments of mass $A$ are incident on F1. Here we assume that, in an ideal binary experiment, a symmetric mass distribution with equal yields of complementary fragments is obtained. $p_\gamma$ is the probability of detecting of a gamma-ray. The quantity $p_\gamma$ is deduced from the response matrix for the HPGe detector under experimental conditions for obtaining a weighted average over a spectrum. The response function was determined by Monte Carlo including the experimental geometry, the effects of gamma-rays transmission and scattering in absorbing materials, the energy resolution broadening, and the total intrinsic efficiency of the HPGe detector. The reliability of the response function was confirmed by measurements employing several standard gamma-ray sources. The measured and simulated spectra for the HPGe detector are compared in Fig. 3. The threshold and time resolution of the detector is around 50 keV and 4.4 ns, respectively. The yield of gamma-ray spectrum below the detector threshold was estimated by linear extrapolation based on the average value near threshold. The average multiplicity of gamma-rays emitted within ~ 5 ns after fission was determined as a function of fragment mass and total kinetic energy.

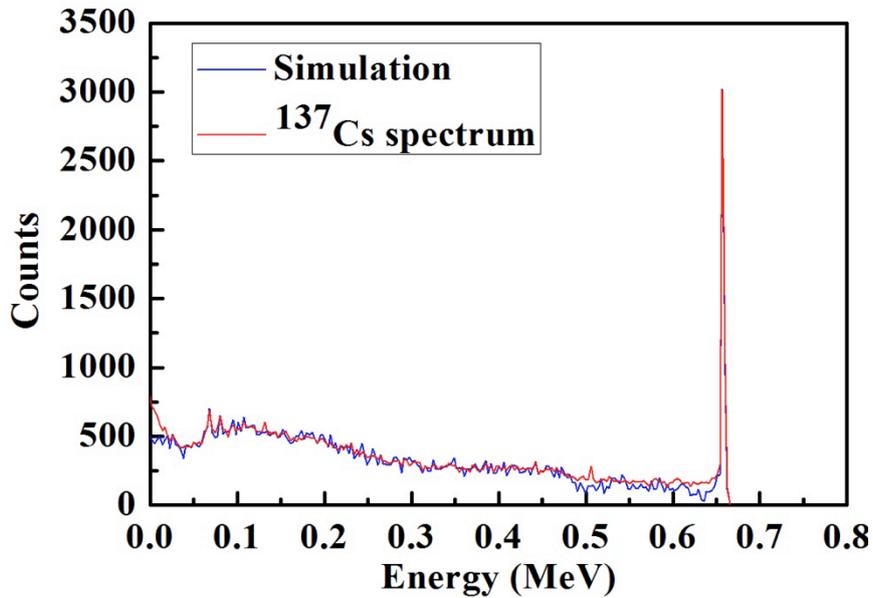

Fig. 3 Comparison of simulated spectra for the HPGe detector and tests with the $^{137}$Cs source.

## 3. Results and discussion

The ratio of the average gamma-ray yield as a function of fragment mass $M_\gamma(A^*)$ to the corresponding average neutron multiplicity $\nu(A^*)$ for individual fragments of $^{252}$Cf(sf), $R(A^*) = M_\gamma(A^*)/\nu(A^*)$, is shown in Fig. 4. The measured ratio $R(A^*)$ is not a strong function of fragment mass. However, there are enhancements at $A_L^* \approx 107$ and $A_H^* \approx 145$. There is a rather pronounced peak near the doubly-magic shell closure of $A^* \approx 132$ where the fragments are stiffer than their nearer neighbors. If prompt gamma-rays originate from vibrational cascades, such a peak may be expected [30] because the average neutron multiplicities are extremely low for fragments near the shell closure ($Z = 50$, $N = 82$) [5], where the gamma-ray enhancement is seen in Fig. 4. See also Ref. [31]. There is also an enhancement for $A^* \approx 85$, near the neutron magic number $N = 50$. The vibrational energy increases as the light fragment become stiffer.

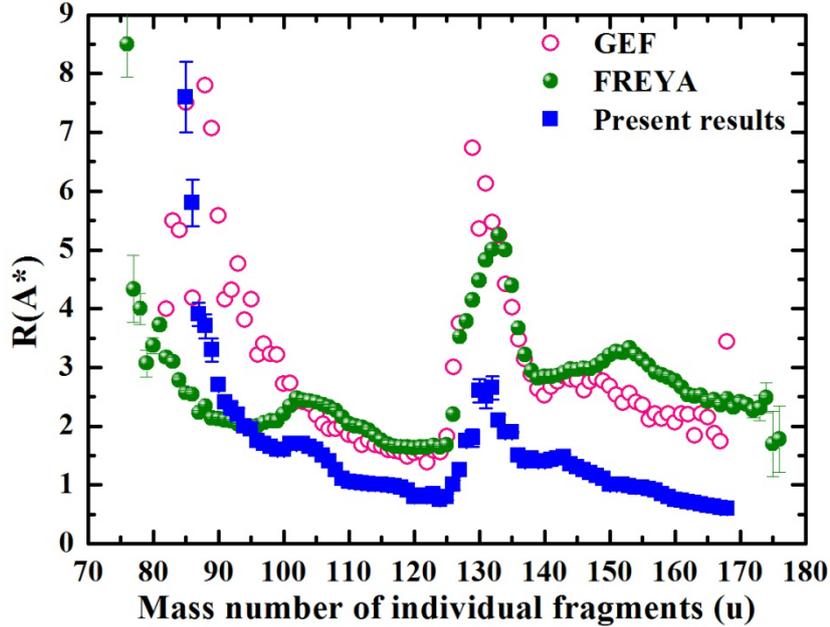

Fig. 4 The dependence of the ratio of average gamma-ray yield $M_\gamma(A^*)$ to the average neutron multiplicity $\nu(A^*)$ on the individual fragment mass.

The nonlinearity of $R(A^*)$ in Fig. 4 indicates that there is no simple relationship between the average neutron multiplicity $\nu(A^*)$ and the average gamma-ray yields $M_\gamma(A^*)$ for $^{252}$Cf(sf). The observed complex relationship can be attributed to the different mechanisms for neutron and gamma-ray emission in fission. The excitation energy of a primary fragment is both collective and statistical. The collective energy is associated with the fragment rotation, while the statistical excitation arises in part from the recovery of the distortion energy at scission. Neutron emission dominates initially and reduces the statistical excitation energy without reducing the rotational energy notably. As this statistical excitation energy approaches the neutron separation energy, the probability for neutron emission decreases and gamma-ray emission takes over [32, 33]. Statistical photon emission, mostly E1 and M1 transitions that carry away relatively little angular momentum, is followed by collective (primarily E2) emission which dissipates the remaining rotational energy and brings the fragments to their ground states.

The ratio $R(A^*)$ has also been calculated with FREYA and GEF. The FREYA result is calculated from a run with 10 million events. The uncertainties on $M_\gamma$ and $\nu$ are added in quadrature. Both FREYA and GEF likely produce more gammas than observed, especially for $A^* > 124$, because both calculations reproduce the trends of $\nu(A^*)$ for $^{252}$Cf(sf). The two models reproduce the general trends of the ratio in shape if not in magnitude.

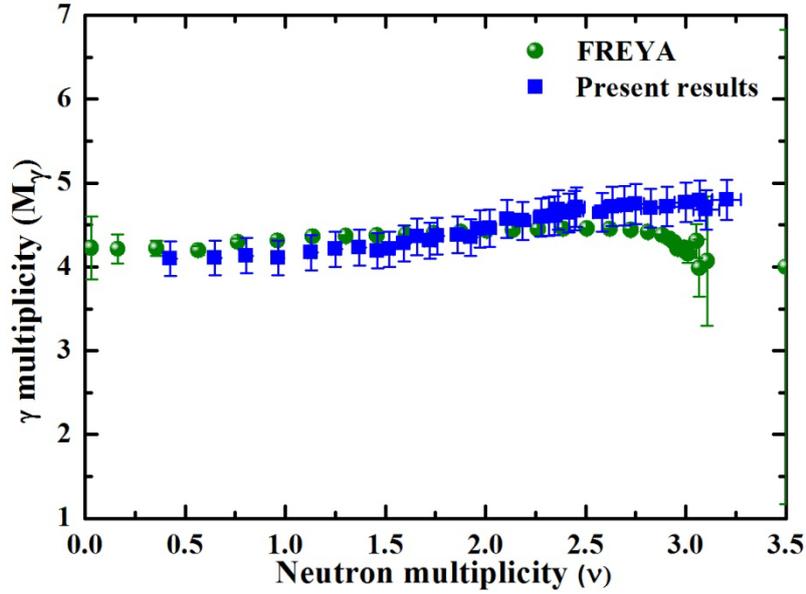

Fig. 5 The correlation of the neutron multiplicity $\nu$ and the gamma-ray multiplicity $M_\gamma$ for light fragments $85 < A^* < 123$.

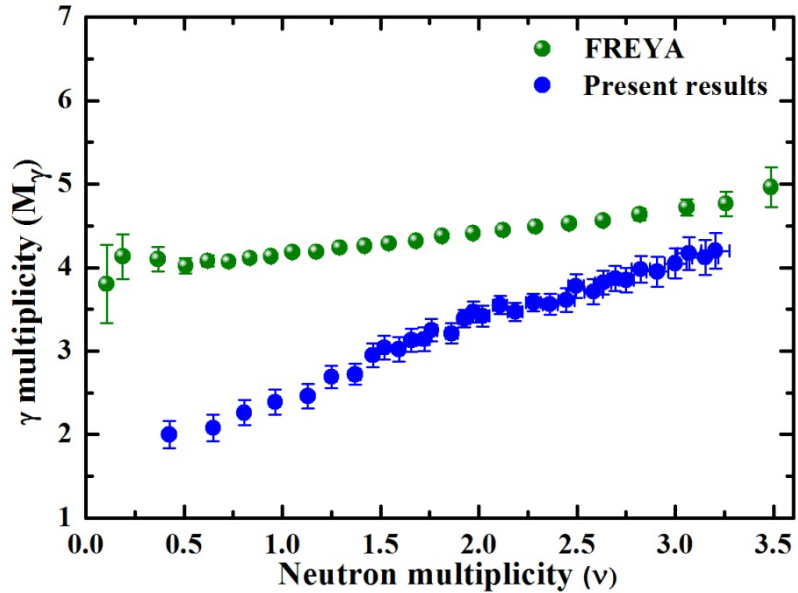

Fig. 6 The correlation of the neutron multiplicity $\nu$ and the gamma-ray multiplicity $M_\gamma$ for symmetric mass fragments $124 < A^* < 131$.

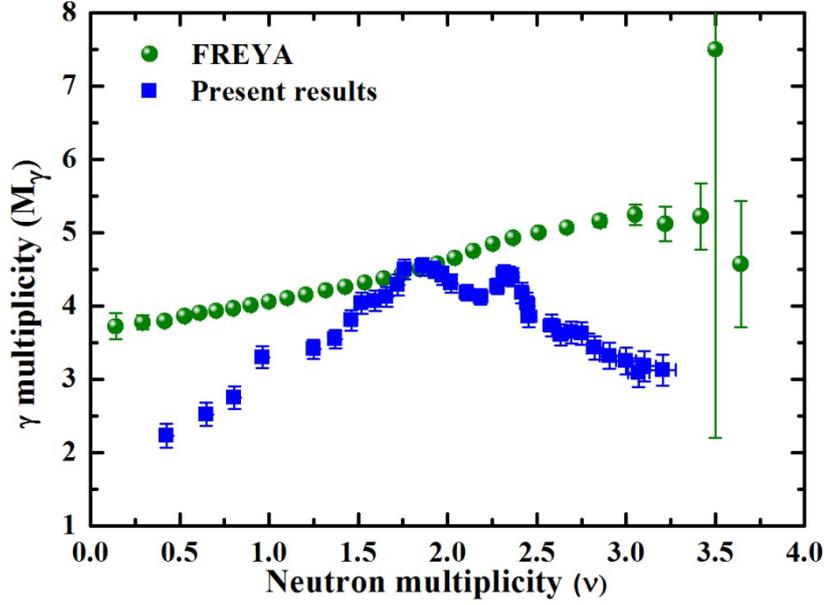

Fig. 7 The correlation of the neutron multiplicity $v$ and the gamma-ray multiplicity $M_\gamma$ for heavy fragments $132 < A^* < 167$.

The dependence of $M_\gamma$ on neutron multiplicity, $M_\gamma(v)$, has been investigated for several different fragment mass regions. For light fragments, $85 < A^* < 123$, $M_\gamma(v)$ has a small positive slope and $<M_{\gamma L}> = 4.48\pm0.23$, as shown in Fig. 5. In the symmetric region, $124 < A^* < 131$, the correlation is remarkably linear with a larger positive slope and $<M_{\gamma S}> = 3.33\pm0.62$, as shown in Fig. 6. In the case of heavy fragments, $132 < A^* < 167$, a stronger and more complex correlation is observed, as shown in Fig. 7. Here $M_\gamma$ rises almost linearly from 2.2 to 4.5 for $0.5 < v < 1.8$, after which it decreases again. There are two rather pronounced peaks at $v = 1.8$ and 2.3. In this case, $<M_{\gamma H}> = 3.78\pm0.59$. Since $<M_{\gamma L}>$ is larger than $<M_{\gamma S}>$ and $<M_{\gamma H}>$, more gamma-rays are emitted from light fragments.

FREYA results are also shown in Figs. 5-7. In each mass region, the neutron and gamma yields are obtained for all masses in TKE bins of 3 MeV. The statistical uncertainties on $M_\gamma$ are shown. For light fragments, the FREYA trend matches the data well. Although the FREYA result is also linear in the symmetric mass region, the positive slope is not as large as that of the data. FREYA also suggests a weak linear dependence of $M_\gamma$ on $v$ for heavy fragment masses and does not exhibit any complex behavior similar to the data. We note that the ground states of the heavy fragments are more likely to be deformed. Thus a larger fraction of gamma emission from the heavy fragment goes through a collective rotational cascade with little statistical variance [9]. This should be taken into account in order to reproduce the behavior shown in the data.

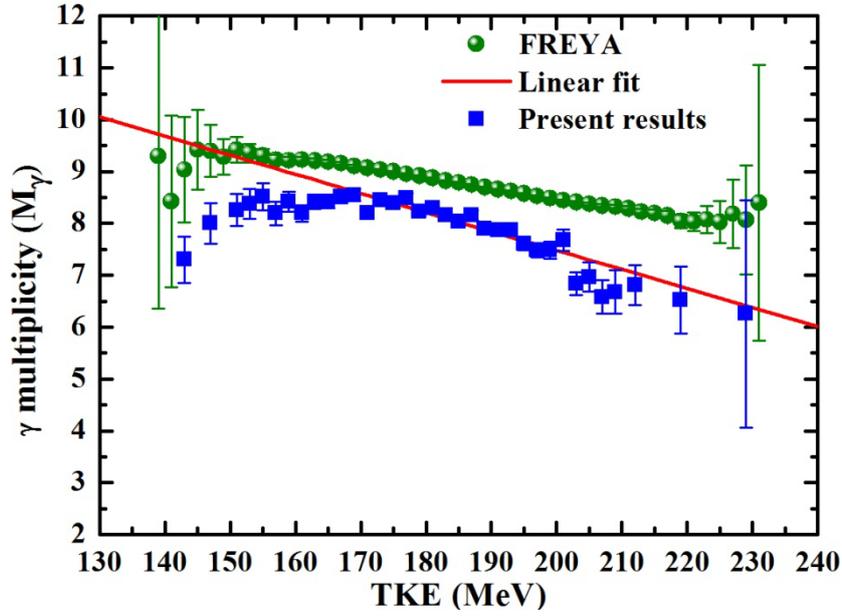

Fig. 8 The gamma-ray multiplicity $M_\gamma$ as a function of the total fragment kinetic energy TKE.

Figure 8 presents the dependence of the average gamma-ray multiplicity on the total fragment kinetic energy, $M_\gamma$(TKE). The measured average multiplicity $M_\gamma$(TKE) is maximal for TKE = 165 – 170 MeV. The data may be fit by a linear function of TKE for TKE > 170 MeV, yielding $d$TKE$/dM_\gamma$ = 27.24 ± 2.84 MeV/gamma. Figure 8 also shows the FREYA results, with statistical uncertainties on the gamma-ray multiplicity. Those results exhibit a behavior similar to the present measurement but with a shallower slope. Below TKE ≈ 170 MeV, the measured $M_\gamma$(TKE) deviates from the linear fit, while there is no significant deviation from a linear behavior in the FREYA result. These data are consistent with our previous measurement [34, 35] where $M_\gamma$(TKE) was obtained using a different detector system consisting of a grid ionization chamber and an NaI(Tl) detector as well as earlier $^{252}$Cf measurements [36] and $^{235}$U($n_{th}$,f) [12]. The variation of $M_\gamma$(TKE) implies that gamma-ray emission is less associated with the initial fragment excitation energy, $E_{ex}^{init} = Q -$ TKE, than neutron emission is, see (TKE) in Ref. [5]. (Here $Q$ is the fission $Q$ value for the given mass partition.) According to Ref. [37], the maximum value of $E_{ex}^{init}$ appears at neither the greatest nor the smallest deformation but for medium nuclear deformations.

The correlation of the total gamma-ray multiplicity TKE is also supported by the results of Krupa *et al*. [17] where a proton beam ($E_p$ = 13, 20, and 55 MeV) was employed in the reactions $^{238}$U($p$,f) and $^{242}$Pu($p$,f). Figure 10 in Ref. [17] shows $M_\gamma$(TKE) for 95 < TKE < 220 MeV in symmetric fission. There is a maximum at TKE ~ 150 MeV. Below this value of TKE, $M_\gamma$(TKE) is monotonically increasing function of TKE while $M_\gamma$(TKE) decreases for TKE > 150 MeV. Similar deviations of the neutron multiplicity from a linear dependence on TKE < 150 MeV in $^{235}$U($n_{th}$, f) [38] and $^{233}$U($n_{th}$, f) [39] as well as TKE < 170 MeV in $^{252}$Cf(sf) [5] have recently been observed. At present, the origin of this behavior is not clear.

The total fragment excitation energy is shared among the intrinsic excitation, nuclear deformation and collective excitation energies [40]. The prompt emission is assumed to be from fully-accelerated fragments. Prompt neutron and gamma emission is followed by delayed emission through beta-decay. While neutron-gamma competition has been considered in many models [20, 41, 42], measured correlations between fission fragments and prompt neutron and gamma emission has been scarce up to now. Many previous gamma measurements were made using NaI detectors that have higher efficiency but poorer energy resolution than the HPGe detector. These neutron and gamma multiplicity measurements are crucial for understanding neutron-gamma competition. The present work helps clarify the relationship between neutron and gamma multiplicity with fragment mass and TKE.

The systematic uncertainties on the results, including detector response and correction factors due to the effects of neutron and gamma-ray emission were calculated by a Monte Carlo. Nuclear data used to obtain these corrections were taken from the ENDF/B-VII.0 evaluation [43] with errors on the order of ~4%. The contributions of extrapolation of gamma-ray pulse height spectrum below the discrimination level is around 0.8%. The count rates of the gamma-ray spectrum and neutron spectrum were low enough for the dead time correction to be ignored. The total uncertainties were calculated from statistical errors together with the systematic errors mentioned above.

## 4. Conclusion

In summary, we have carried out a new investigation of the dependence of the relationship between the average neutron multiplicity $v$ and the average gamma-ray multiplicity $M_\gamma$ in spontaneous fission of $^{252}$Cf as functions of the fragment mass $A^*$ and total kinetic energy TKE, based on the ratio of $M_\gamma/v$ and the function $M_\gamma(v)$. For the first time a positive correlation was shown to exist between the gamma-ray yield $M_\gamma$ and the neutron multiplicity $v$ in the light and symmetric fragment mass regions, while a complex relationship was observed in the heavy fragment mass region. This relationship cannot be understood in terms of current complete event fission models. The ratio $M_\gamma/v$ exhibits strong shell effects near neutron magic number $N = 50$ and near the doubly-magic nucleus $(Z,N) = (50,82)$. The gamma-ray multiplicity $M_\gamma$ is a maximum for TKE=165-170 MeV and decreases linearly for TKE >170 MeV. Below TKE≈170 MeV, $M_\gamma$ deviates from the linear fit. These detailed experimental results can not only provide new information on the correlations but also reveal any anomalies in neutron and gamma-ray emission in spontaneous fission of $^{252}$Cf.

**Acknowledgements**

This work has been supported by the National Natural Science Foundation of China (No. 10175091, No. 11305007 and No.11235002). The work of Ramona Vogt and Jørgen Randrup was supported by the Office of Nuclear Physics in the U.S. Department of Energy's Office of Science under Contracts No.DE-AC02-05CH11231 (JR) and DE-AC52-07NA27344 (RV). We also gratefully acknowledge support of